\documentclass[prb,twocolumn,showpacs,superscriptaddress,amsmath,floatfix]{revtex4}
\usepackage{graphicx}
\usepackage{amssymb}
\usepackage{citesort}

\begin{document}
\title{Monte Carlo simulations of two-dimensional fermion systems with string-bond states}
\author{J. -P. Song}
\affiliation{Department of Physics and Astronomy and HPC$^2$ Center for 
Computational Sciences, Mississippi State University, Mississippi State MS 39762}
\author{R.T. Clay}
\affiliation{Department of Physics and Astronomy and HPC$^2$ Center for 
Computational Sciences, Mississippi State University, Mississippi State MS 39762}
\date{\today}
\begin{abstract}
We describe an application of variational Monte Carlo to
two-dimensional fermionic systems within the recently developed
tensor-network string-bond state (SBS) ansatz.  We use a combination
of variational Monte Carlo and stochastic optimization to optimize the
matrix-product state matrices representing the ground state.  We
present results for a two-dimensional spinless fermion model including
nearest-neighbor Coulomb interactions and determine using finite-size
scaling the phase boundary between charge-ordered insulating and
metallic phases.  This approach can treat frustrated systems and be
easily extended to for fermions with spin.
\end{abstract}

\pacs{02.70.Ss, 71.10.Fd, 71.10.Hf}
\maketitle

\section{Introduction}
\label{intro}

The properties of two dimensional (2D) and frustrated quantum
many-body models play an important role in condensed matter physics.
Numerical methods including quantum Monte Carlo (QMC) \cite{Evertz03a}
and the density matrix renormalization group (DMRG)
\cite{White92a,White93a,Schollwock05a} have been essential in
understanding the ground state and thermodynamic properties of
interacting electron and spin systems. These two classes of methods
have well known limitations however: QMC is severely limited to the
systems that can be studied by the fermion sign problem, and DMRG
methods are largely limited to one dimensional (1D) or quasi-1D
systems.

Underlying DMRG methods is a matrix product state (MPS) representation
of the quantum state. If each configuration in the wavefunction is
written as $|s_1,\ldots, s_N\rangle$ where $s_i$ denote local quantum
degrees of freedom such as the spin $S^z_i$ on the $i$-th lattice site
and $N$ is the total number of sites in the lattice,
a MPS representation for the wavefunction $|\Psi\rangle$ is written as
\begin{equation}
|\Psi\rangle=\sum_{s_1,\ldots,s_N} {\rm Tr} 
\left[A^1_{s_1}\cdots A^N_{s_N}\right]
|s_1,\ldots,s_N\rangle.
\label{eq-mps}
\end{equation}
In Eq.~\ref{eq-mps} the weight of each configuration is given by the
trace of a product of $D\times D$ matrices $A^i_{jk}$ The advantage of
using a MPS representation is that provides an accurate representation
of the ground state of a 1D quantum system with only
moderate\cite{Rommer97a,Schollwock11a} values of $D$.
Eq.~\ref{eq-mps} can be used to represent a 2D system by simply
numbering the lattice sites in 2D sequentially (as in
Fig.~\ref{fig-contractions}(a)), but favorable scaling with the matrix
size $D$ is then lost because the MPS ansatz can only describe
entanglement along one chain direction.

A recent innovation is the use of Monte Carlo sampling to evaluate
expectation values of the Hamiltonian as well as other operators
within MPS-type trial states
\cite{Sandvik07a,Schuch08a,Sfondrini10a,Wang11b}.  By sampling the
physical states of the system rather than contracting the matrices the
computational scaling in $D$ is reduced.  Derivatives of the energy
with respect to the matrix elements can also be calculated and then
used to optimize the matrix elements $A^i_{jk}$
\cite{Sandvik07a,Sfondrini10a}. The use of QMC sampling brings the
computational advantage of trivial parallelization of Monte Carlo
averages. While most applications have been to quantum spin models,
this approach has successfully been used for more complicated quantum
models such as the 1D Hubbard model where each site has four rather
than two degrees of freedom \cite{Clay12a}.

Many variations of the MPS ansatz have been suggested to generalize it
to 2D systems.  The most natural extension to higher dimensions is to
replace the matrices in Eq.~\ref{eq-mps} by tensors and the trace by a
more general contraction over the tensor indices. Projected entangled
pair states (PEPS) are one such tensor network generalization
\cite{Verstraete04a}.  PEPS have been successfully applied to 2D
frustrated spin models \cite{Murg07a,Schuch08a,Sfondrini10a}.  A
variation (iPEPS) has also been proposed for evaluating thermodynamic
(infinite lattice) quantities \cite{Jordan08a,Orus09a,Vidal07a}.  The
main limitation in applying these methods is their poor computational
scaling in the tensor size\cite{Murg07a,Verstraete08a}, typically
$\propto D^{12}$.  An alternate approach is to use a somewhat more
restricted ansatz that can be more easily computationally
evaluated. The promise is that one can trade some complexity of the
representation by increasing the number of variational
parameters. Examples of this general approach include the multi-scale
entanglement renormalization ansatz (MERA) \cite{Vidal07b}, second
renormalization of tensor networks \cite{Xie09a,Zhao10a}, and
tensor-renormalization group \cite{Levin07a,Wang11a} approaches. In
this paper we will explore a generalization of one such approach, the
string-bond states (SBS) ansatz, where several one dimensional MPS
``strings'' of operators are placed in different directions on the 2D
lattice \cite{Schuch08a,Sfondrini10a}.

Applications to fermionic systems bring additional challenges to
tensor network methods. 
In an occupation number representation the sign of
each configuration necessarily depends on the ordering of the
fermionic creation operators. While trivial in 1D where the
Jordan-Wigner transformation can be used, the signs lead to
long-range correlations between the tensors representing neighboring
sites in 2D.  One of the key questions is whether it is
possible to come up with an effectively local tensor network scheme
for fermions and to what degree the difficulty of doing this depends
on the model in question.  Several approaches have nevertheless been
proposed to treat fermionic systems by making the required tensor
operations local.  Examples include the modification of MERA by
introducing fermionic ``swap gates'' \cite{Corboz09a,Corboz10b} and
fermionic versions of PEPS
\cite{Corboz10a,Corboz10c,Kraus10a,Pizorn10a}.  The principal
disadvantage to using these methods for practical calculations is
again that while they scale as a polynomial in tensor bond dimension,
the polynomial power is typically large. The approach we present here
is to simply keep the long range correlations between local matrices
$A^i_{jk}$ and use optimization to find the required signs.  The
advantage is that the formal scaling in matrix size $D$ remains small so
much larger $D$ can be reached.  The disadvantage is that this potentially
 leads to a more challenging optimization problem. However, we
will show that in practice good results for 2D fermionic systems on
significant lattice sizes (up to 12$\times$12) can be reached
within the SBS approximation. 

The paper is organized as follows. Details of our SBS-QMC method are
discussed in Section~\ref{method}. In Section~\ref{results} we show
results for a 2D interacting spinless fermion system, followed by
further discussion in Section~\ref{discussion}.

\section{Method}
\label{method}

\begin{figure}[tb]
\centerline{\resizebox{3.0in}{!}{\includegraphics{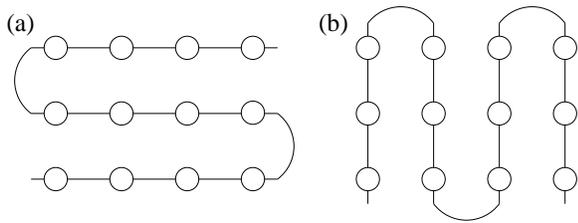}}}
\caption{Contraction patterns for (a) string $S_A$ composed of $A$ matrices and (b) string
$S_B$ of $B$ matrices, illustrated
for a 4$\times$3 lattice.}
\label{fig-contractions}
\end{figure}

For a generic Hubbard-type model we decompose the Hamiltonian into two terms,
\begin{equation}
H=H_0+H_1,
\end{equation}
where diagonal $H_0$ and off-diagonal $H_1$ terms are given by
\begin{eqnarray}
H_0&=&U\sum_{i}n_{i\uparrow}n_{i\downarrow}
+\sum_{\langle i,j \rangle} V_{ij} n_i n_j, \nonumber \\
H_1&=&-\sum_{\langle i,j \rangle \sigma} t_{ij}
(c^\dag_{j\sigma}c_{i\sigma}+
c^\dag_{i\sigma}c_{j\sigma}).
\label{eq-hubbard}
\end{eqnarray}
In Eq.~\ref{eq-hubbard}, $c^\dag_{i\sigma}$ ($c_{i\sigma}$) create
(annihilate) an electron of spin $\sigma$ on site $i$,
$n_{i\sigma}=c^\dagger_{i\sigma}c_{i\sigma}$, and $n_i=n_{i\uparrow}+n_{i\downarrow}$.
We assume here that the nearest neighbor sites in $H_1$ are those
on a conventional square lattice, although as discussed later,
it is possible to generalize this to other periodic lattices.
$U$ and $V_{ij}$ are on-site and intersite Coulomb interactions.
The weight of a configuration in the SBS approximation is
 represented in terms of overlaps
defined on a set of operator ``strings'' $\{S\}$:
\begin{equation}
\langle C_n |\Psi\rangle=\prod_{S} {\rm Tr} \left[\prod_{i} S^i \right],
\end{equation}
where $|C_n\rangle$ is a state in a local (e.g. occupation number) basis
and $S^i$ are $D\times D$ matrices.  As shown in
Fig.~\ref{fig-contractions} we use a set of two strings $\{S_A,S_B\}$
to cover the lattice, each of which corresponds to the usual ``snake''
generalization conventionally used to adapt a MPS state to a 2D geometry.
The string $S_A$ ($S_B$) follows the hopping integrals aligned
along $x$ $(y)$.  The matrices for these two strings are labeled $A$
and $B$.
The SBS representation for the wavefunction $|\Psi\rangle$ is written as
\begin{equation}
|\Psi\rangle=\sum_{n}W(C_n))|C_n\rangle,
\label{eq-sbs}
\end{equation}
where $W(C_n)=\prod_S W_S(C_n)$.
The weights $W_S(C_n)$ for the two strings are given by
\begin{eqnarray}
W_A(C_n)={\rm Tr}\prod_{i\in S_A}^N A^i
={\rm Tr}\prod_{i_x =1}^L \left( \prod_{i_y =1}^M A^{M(i_x-1)+i_y} \right), \\
W_B(C_n)={\rm Tr}\prod_{i\in S_B}^N B^i
={\rm Tr}\prod_{i_y =1}^M \left( \prod_{i_x =1}^L B^{M(i_x-1)+i_y} \right),
\label{eq-weights}
\end{eqnarray}
where $i_x$ and $i_y$ correspond to the $x$ and $y$ coordinates of site
$i=(i_x,i_y)$ for a $L\times M$ rectangular lattice with the total number of
lattice sites $N$.

The variational Monte Carlo (MC) method we use to evaluate the energy
and other correlation functions is based on the method of reference
\onlinecite{Sandvik07a}.  We have previously shown that this method
can be generalized to 1D fermionic systems \cite{Clay12a}, where the
weight of a configuration is given by a MPS, i.e. a single string.
Configurations $|C_n\rangle$ are sampled according to the weight
$W(C_n)^2$.  MC updates consist of interchanges of electrons of a
given spin between neighboring sites. Updates are attempted first
along the path of string $S_A$ and then along the direction of string
$S_B$. In this manner, a system of ``left'' and ``right'' matrices can
be used to efficiently perform the MC sampling \cite{Sandvik07a}.  We
create a series of left matrices $L_A^{i_x,i_y}=A^iL_A^{i_x,i_y+1}$
and $L_B^{i_y,i_x}=B^iL_B^{i_y,i_x+1}$ for $i_x=1,\ldots L$ and
$i_y=1,\ldots M$.  Sequentially visiting the site $i=(i_x,i_y)$ in
either horizontal $x$ (for $S_A$) or vertical $y$ (for $S_B$)
direction, we attempt to interchange electrons between that site and
its nearest neighbor $j=(j_x,j_y)$ until we have arrived at site
$N=(L,M)$. If a update is accepted (or rejected) according an
acceptance probability $p(C_{n} \rightarrow C_{n\prime})=\min
[W^2(C_{n\prime})/W^2(C_n),1]$, the right matrices
$R_A^{i_x,i_y}=R_A^{i_x,i_y-1}A^i$ and
$R_B^{i_y,i_x}=R_B^{i_y,i_x-1}B^i$ are advanced, respectively.  Once
the $R$ matrices for a given string have been stored, measurements of
the energy and derivatives of the energy are are implemented by
traversing the string in the reverse direction \cite{Sandvik07a}.

The energy estimator for the configuration $C_n$ is
\begin{equation}
E(C_n)=\sum_{C_{n'}} \frac{W(C_{n'})}{W(C_n)}
	     \langle C_{n'} |H|C_n\rangle.
\label{eq-estimator}
\end{equation}
In Eq.~\ref{eq-estimator}, the diagonal part of the energy $\langle
H_0 \rangle$ can simply be measured as an average over the
configurations visited.  Interchanges of electrons give
contributions to the off-diagonal terms 
$\langle H_1\rangle$.  In calculating the matrix
element in Eq.~\ref{eq-estimator} a sign due to fermion exchange must
be included. 

Within the MPS representation the derivative of the energy with respect to the
each of the matrix elements can easily be calculated. For the $A$
matrices of $S_A$  this derivative is
\begin{equation}
\frac{\partial E}{\partial A^k_{ij}} = 2 \left\langle
\frac{E(C_n)- \big\langle E(C_n) \big\rangle}{W_A(C_n)}
\frac{\partial W_A(C_n)}{\partial A^k_{ij}}
\right\rangle, 
\label{eq-energyderiv}
\end{equation}
where the derivatives of each trace can be written as
\begin{equation}
\frac{\partial W_A(C_n)}{\partial A^k_{ij}}
= \frac{1}{1+\delta_{ij}}
\left[ Q^A_{ij}(k) + Q^A_{ji}(k) \right],
\label{eq-weightderiv}
\end{equation}
using $Q^A(k)=\prod_{i\neq k}A^i$.
An identical expression is used for derivatives of the energy
with respect to the $B$ matrices

The matrix elements $A^k_{ij}$ and $B^k_{ij}$ for $k=1,\ldots,N$ 
are first initialized to
random numbers in the interval $[-\frac{1}{2},\frac{1}{2}]$.  We
normalize the matrices so that their Frobenius norm is unity,
i.e. $\frac{1}{D}{\rm Tr}(AA^T)=1$.  MC measurements for the energy,
derivatives, and other correlation functions are block-averaged as
usual. 
After each block, matrix elements are updated using a 
stochastic optimization scheme \cite{Sandvik07a}. 
Each matrix element $A^k_{ij}$ is modified by a random
amount in the direction indicated by the derivative of the energy,
\begin{equation}
A^k_{ij} \rightarrow A^k_{ij}
-\delta \cdot R\cdot {\rm sgn} \left(\frac{\partial E}{\partial A^k_{ij}}\right)
\theta\left(\left|\frac{\partial E}{\partial A^k_{ij}}\right|-\alpha\right).
\label{eq-matrixupdate}
\end{equation}
Here $R$ is a random number in the interval $[0,1)$,
${\rm sgn}(x)$ is the signum function of a real number $x$,
and $\theta(x)$ is the unit step function.
The parameter  $\delta$ sets the maximum change for a matrix element.
The parameter $\alpha$ restricts changes to only the matrix elements
that have the most significant effect on the energy, those with the
largest magnitude derivatives.
We found that a small reduction in energy with a suitable choice of  
alpha. The reduction in the energy was small compared to the standard  
stochastic optimization, but it improves convergence. This effect help  
to reduce unwanted stochastic noise as the global minimum is reached,  
therefore, a lower energy can be obtained.
  Several MC blocks each followed by the update in
Eq.~\ref{eq-matrixupdate} are then combined into one {\it step}
labeled by the index $k$ of the optimization algorithm (see
Fig.~\ref{fig-energy}). At each successive $k$ the parameters $\delta$
and $\alpha$ are decreased by a multiplicative factor $Q$. For the
results here, we typically used
$Q=0.9$. $\delta$ and $\alpha$ were initially chosen as 0.5.
 Simultaneously the number of MC blocks per
step, $G(k)$, and samples per block, $F(k)$, are increased linearly,
We typically used $F(k)=$5000-10000 and $G(k)=$250-500.
This procedure gives an ``annealing'' procedure that for a sufficiently
large $k$ should approach the global minimum energy.
The MC sampling was parallelized using an ``embarrassingly parallel''
algorithm. The results presented here used up to 192 processors.

\section{Results}
\label{results}

\begin{figure}[tb]
\centerline{\resizebox{3.0in}{!}{\includegraphics{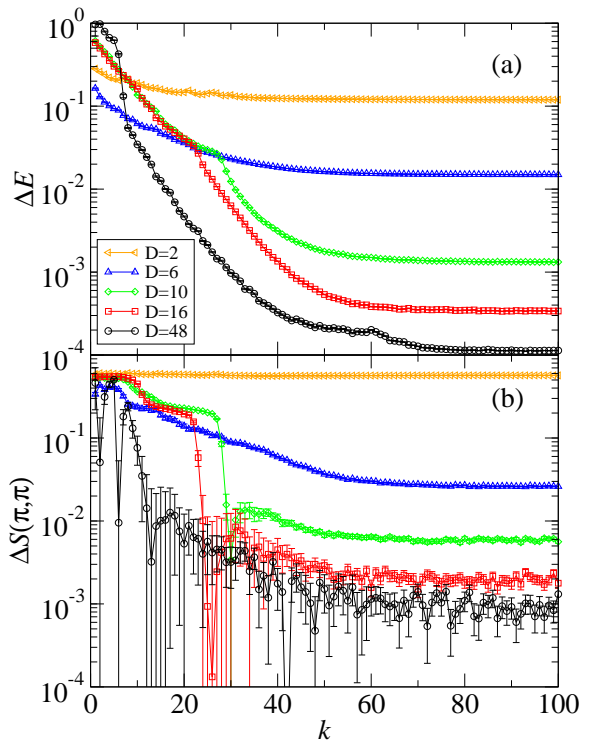}}}
\caption{(color online) Relative error of the (a) ground state energy
  and (b) charge structure factor at $S(\pi,\pi)$, as a function
  of the number of algorithm steps $k$ (see text) and matrix size $D$
  for a 4$\times$4 periodic lattice with 16 particles and $V/t=0.45$.}
\label{fig-energy}
\end{figure}

We consider spinless fermions on a 2D square
lattice interacting with a nearest-neighbor Coulomb repulsion.
The Hamiltonian is given by
\begin{equation}
H=-t\sum_{\langle ij\rangle}(c^\dagger_ic_j+H.c.) + V\sum_{ij} n_in_j.
\label{ham}
\end{equation}
In Eq.~\ref{ham}, $c^\dagger_i$ creates a fermion on site $i$; sites
$i$ and $j$ in $\langle ij \rangle$ are nearest-neighbor pairs on a 2D
square lattice of $N$ sites with periodic boundary conditions.
 All energies will be given in units of $t$.  We 
consider the half-filled case with $N/2$ particles. For this density,
the $V$ interaction causes a checkerboard pattern charge-ordered (CO)
insulating phase.  In the 1D limit the model may be transformed via
the Jordan-Wigner transformation to a spin-$\frac{1}{2}$ XXZ
Heisenberg model and it can be shown exactly that the CO phase occurs
when $V>V_c$ with\cite{Mila93a} $V_c=2$.  In 2D $V_c$ is not known
exactly.  Analytical work using a slave-boson approximation was done
for a model with SU(N) fermions \cite{McKenzie01a}. For the case of a
2D square lattice and taking $N=2$ (corresponding to
spin-$\frac{1}{2}$) the corresponding $V_c=0.69$.
\begin{figure}[tb]
\centerline{\resizebox{3.0in}{!}{\includegraphics{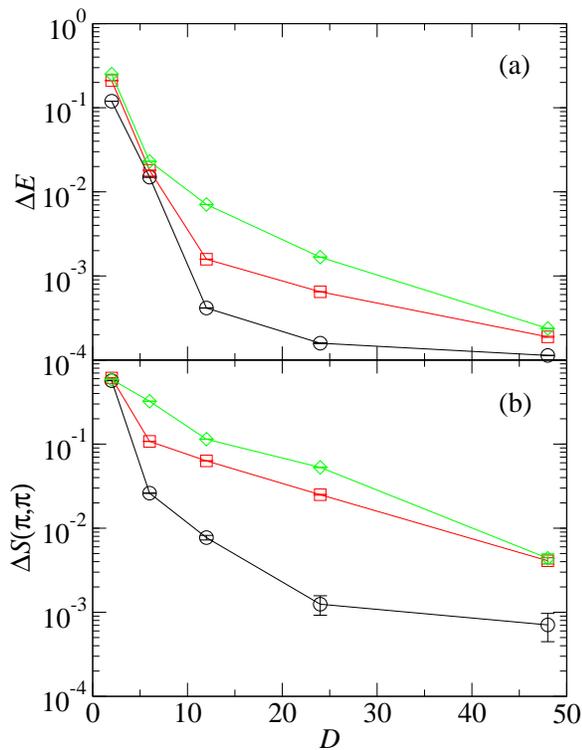}}}
\caption{(color online) (a) Relative error of the ground state energy
  and (b) of the charge structure factor as a function of matrix size
  $D$ for $V/t=0.45$.  Circles, squares and diamonds are for
  4$\times$4, 6$\times$4 and 8$\times$4 lattices, respectively.}
\label{fig-energy-charge}
\end{figure}
This model was also previously  studied using finite-temperature
determinantal QMC\cite{Gubernatis85a} down to temperatures of order
$T\sim 0.5$. These numerical results were also compared with the
mean-field RPA predictions \cite{Gubernatis85a}.  If one extrapolates
the strong-coupling RPA result from reference
\onlinecite{Gubernatis85a} to $T=0$, $V_c\approx 1/\sqrt{3} \approx
0.58$. The finite-temperature QMC results for $V_c$ appear to be
consistent with this limit if an almost-linear extrapolation in the
$T-V$ plane is assumed, but could not rule out the possibility that
$V_c\rightarrow 0$ as $T\rightarrow 0$. As shown below, our present results
are consistent with a nonzero $V_c$.

We compared the SBS-QMC results to exact diagonalization calculations
for systems up to 32 sites.  Fig.~\ref{fig-energy}(a) shows the
relative error in the ground state energy, $\Delta
E=|(E_{\rm{QMC}}-E_{\rm exact})/E_{\rm exact}|$, as a function of
algorithm steps $k$ and matrix size $D$ for a 4$\times$4 lattice.  The
interaction strength $V=0.45$ chosen here is close to the CO
transition point representing the most computationally challenging
parameter region of the model.  Here and in our following results,
each value of $D$ is a separate calculation, each starting with
different random initial matrix elements.  In comparison with quasi-1D
systems where a single MPS can be used to represent the wavefunction
(see Fig.~A-1 of Reference \onlinecite{Clay12a}), we found nearly
comparable scaling of accuracy with respect to $D$ for the 2D system
considered here.

An order parameter for the CO phase is the charge structure
factor $S(\boldsymbol{q})$ for $\boldsymbol{q}=(\pi,\pi)$, where $S(\boldsymbol{q})$
is defined as
\begin{equation}
S(\boldsymbol{q})=\frac{1}{N}\sum_{j,k} e^{i\boldsymbol{q} \cdot \boldsymbol{r}_{jk}}
\langle (n_j-\frac{1}{2})(n_k-\frac{1}{2}) \rangle.
\end{equation}
Fig.~\ref{fig-energy}(b) shows the
relative error in  $S$, $\Delta S=|(S_{\rm{QMC}}-S_{\rm{exact}})/S_{\rm{exact}}|$ for a $4\times4$ lattice
at the ordering wavevector $\boldsymbol{q}=(\pi,\pi)$.

\begin{figure}[tb]
\centerline{\resizebox{3.0in}{!}{\includegraphics{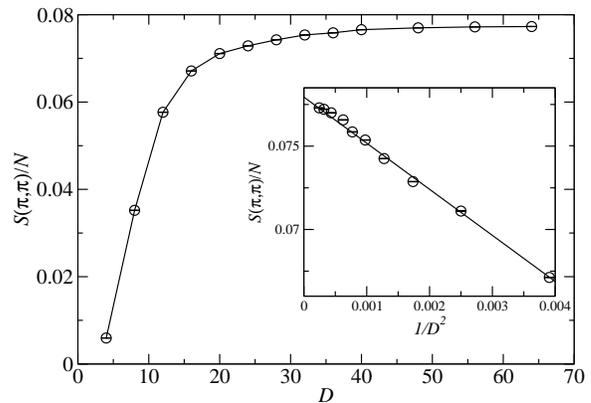}}}
\caption{
 The charge structure factor $S(\pi,\pi)/N$
as a function of matrix size $D$ on a 12$\times$12 for $V/t=0.8$. The
inset shows the same data plotted versus $1/D^2$.}
\label{fig-sfactor}
\end{figure}

Fig.~\ref{fig-energy-charge}(a) and (b) further  show the relative errors as a function
of matrix size $D$ for larger system sizes that can still be solved exactly.
 As expected and seen in Fig.~\ref{fig-energy-charge}, for larger systems larger values of
$D$ are required to reach the same accuracy.  In all of the
comparisons in Figs.~\ref{fig-energy} and \ref{fig-energy-charge}, of
order 100 algorithm steps were needed to converge the energy to within
a relative energy accuracy of order $10^{-4}$. We also verified that
restarting the optimization from the converged matrices gave no
further improvement in the energy.
Fig.~\ref{fig-sfactor} shows the convergence with $D$ for the largest
system studied, 12$\times$12, with $V=0.8$.  As shown in the inset,
$S(\pi,\pi)$ scales as approximately $1/D^2$. 
\begin{figure}[tb]
\centerline{\resizebox{3.0in}{!}{\includegraphics{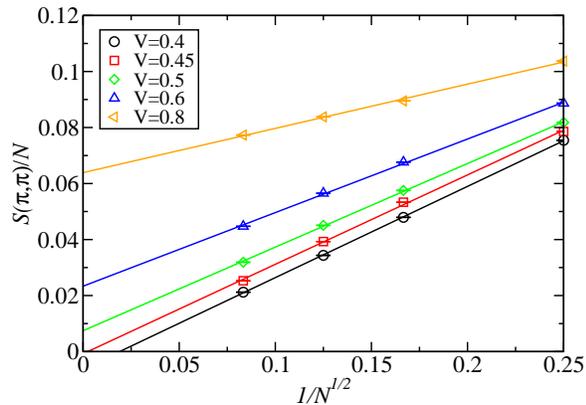}}}
\caption{(color online) Finite-size scaling of the charge structure
  factor $S(\pi,\pi)/N$ versus $1/N^{1/2}$ for spinless fermions on
  square periodic lattices at half-filling.  SBS-QMC simulations were
  performed for up to 12$\times$12 systems; $S(\pi,\pi)$ for the
  smallest system size was calculated exactly.}
\label{fig-scaling}
\end{figure}

In the CO phase $S(\pi,\pi)/N$ converges to a finite value in the
thermodynamic limit.  Fig.~\ref{fig-scaling} shows the finite-size
scaling of $S(\pi,\pi)/N$.  The results in Fig.~\ref{fig-scaling} used
up to $D=64$ matrices and clearly show that a finite critical coupling
$V_c$ for the CO phase exists.  By plotting the extrapolated
$S(\pi,\pi)/N$ versus $V$, we estimate that $V_c$ for the CO
transition is $V_c=0.45\pm0.02$

\section{Discussion}
\label{discussion}

In this paper we have presented numerical results using the SBS ansatz
applied to a 2D fermionic model. In order to simulate a fermionic
system, we have not attempted to make the sign pattern local, but
instead have simply used stochastic optimization to optimize both the
sign and amplitude of a general SBS wavefunction. Because the
computational scaling of the method is relatively small
(proportional\cite{Sandvik07a} to $ND^3$), this ``brute force''
optimization is successful for reasonably large fermionic systems, for
example here up to $N=144$. As the method is not restricted to
unfrustrated lattices, we expect it will provide a useful way to study
frustrated Hubbard-type models on lattice sizes out of reach of exact
diagonalization. In comparison with DMRG which is more accurate on
rectangular lattices of large aspect ratio, the SBS-QMC method can be
used on square periodic lattices which are the easiest to perform
finite-size scaling on.

While we have presented data here for a spinless fermion model, we are
presently testing the method for 2D frustrated models including spin.
Incorporating spin simply increases the number of states per site,
which we find requires a somewhat larger $D$ to obtain comparable
accuracy in the energy and correlation functions.  Further
improvements on the algorithm also can certainly be made. In applying SBS
to 2D spin systems, it was noticed that the initial choice for the
matrix elements could make a large difference in the convergence
\cite{Sfondrini10a}. Here we have only used random starting
matrices--using a mean-field solution as the initial starting state
could potentially improve the results.

\section{Acknowledgments}

This work was supported by the US Department of Energy grant
DE-FG02-06ER46315.  We thank A. Sandvik for helpful discussions while
preparing this manuscript. RTC thanks the Condensed Matter Theory
Visitor's Program at Boston University for hospitality while on sabbatical.

\end{document}